\documentclass[sigconf, nonacm]{acmart}

\usepackage{xcolor}
\usepackage{enumitem}
\usepackage{amsmath}
\usepackage{dblfloatfix}
\usepackage{cleveref}

\usepackage{subfiles}
\usepackage{blindtext} 
\usepackage{listings}
\usepackage{array}
\usepackage{soul}
\usepackage{xspace}
\usepackage[ruled,vlined,linesnumbered]{algorithm2e}

\usepackage{colortbl}
\usepackage{booktabs}
\usepackage{tikz}
\usetikzlibrary{positioning, shapes.multipart, calc, arrows.meta}
\usepackage{subcaption}
\usepackage{graphicx}
\usepackage{balance}
\usepackage{tabularx}
\graphicspath{{images/}{../images/}}
\usepackage{adjustbox}

\usepackage{pgf}
\usepackage{tikz}
\usetikzlibrary{patterns}
\usetikzlibrary{decorations.pathmorphing} 

\definecolor{brewerpurple}{HTML}{AF4EA3}

\definecolor{brewerlightblue}{HTML}{a6cee3}
\definecolor{brewerblue}{HTML}{1f78b4}
\definecolor{brewergreen}{HTML}{b2df8a}

\definecolor{linegray}{gray}{0.5} 

\lstdefinestyle{custompython}{
    language=Python,
    basicstyle=\ttfamily\small,
    keywordstyle=\color{brewerblue},
    morekeywords={with},
    commentstyle=\color{gray},
    stringstyle=\color{brewergreen},
    numbers=left,
    numberstyle=\tiny\color{linegray},
    numbersep=8pt,
    emph={flor,torch,random},emphstyle=\bfseries\color{brewerpurple},
}

\lstdefinestyle{sqlc}{
    mathescape=true,
    language=SQL,
    showspaces=false,
    showstringspaces=false,
    basicstyle=\ttfamily,
    commentstyle=\color{gray},
    morekeywords={crosstab, with}
}

\lstdefinestyle{PythonPandas}{
    language=Python,
    basicstyle=\ttfamily\small,
    keywordstyle=\color[rgb]{1,0.4,0},
    stringstyle=\color[rgb]{0,0.5,0},
    commentstyle=\color[rgb]{0,0.5,1},
    numberstyle=\tiny\color[rgb]{0.5,0,0.5},
    stepnumber=1,
    numbersep=10pt,
    tabsize=4,
    showspaces=false,
    showstringspaces=false
}

\definecolor{gitgreen}{RGB}{0, 128, 0}

\makeatletter
\pgfarrowsdeclare{crow's foot}{crow's foot}
{
    \pgfarrowsleftextend{+-.5\pgflinewidth}%
    \pgfarrowsrightextend{+.5\pgflinewidth}%
}
{
    \pgfutil@tempdima=0.6pt%
    \pgfsetdash{}{+0pt}%
    \pgfsetmiterjoin%
    \pgfpathmoveto{\pgfqpoint{0pt}{-9\pgfutil@tempdima}}%
    \pgfpathlineto{\pgfqpoint{-13\pgfutil@tempdima}{0pt}}%
    \pgfpathlineto{\pgfqpoint{0pt}{9\pgfutil@tempdima}}%
    \pgfpathmoveto{\pgfqpoint{0\pgfutil@tempdima}{0\pgfutil@tempdima}}%
    \pgfpathmoveto{\pgfqpoint{-8pt}{-6pt}}%
    \pgfpathlineto{\pgfqpoint{-8pt}{-6pt}}%
    \pgfpathlineto{\pgfqpoint{-8pt}{6pt}}%
    \pgfusepathqstroke%
}

\pgfarrowsdeclare{omany}{omany}
{
    \pgfarrowsleftextend{+-.5\pgflinewidth}%
    \pgfarrowsrightextend{+.5\pgflinewidth}%
}
{
    \pgfutil@tempdima=0.6pt%
    \pgfsetdash{}{+0pt}%
    \pgfsetmiterjoin%
    \pgfpathmoveto{\pgfqpoint{0pt}{-9\pgfutil@tempdima}}%
    \pgfpathlineto{\pgfqpoint{-13\pgfutil@tempdima}{0pt}}%
    \pgfpathlineto{\pgfqpoint{0pt}{9\pgfutil@tempdima}}%
    \pgfpathmoveto{\pgfqpoint{0\pgfutil@tempdima}{0\pgfutil@tempdima}}%
    \pgfpathmoveto{\pgfqpoint{0\pgfutil@tempdima}{0\pgfutil@tempdima}}%
    \pgfpathmoveto{\pgfqpoint{-6pt}{-6pt}}%
    \pgfpathcircle{\pgfpoint{-11.5pt}{0}} {3.5pt}
    \pgfusepathqstroke%
}

\pgfarrowsdeclare{one}{one}
{
    \pgfarrowsleftextend{+-.5\pgflinewidth}%
    \pgfarrowsrightextend{+.5\pgflinewidth}%
}
{
    \pgfutil@tempdima=0.6pt%
    \pgfsetdash{}{+0pt}%
    \pgfsetmiterjoin%
    \pgfpathmoveto{\pgfqpoint{0\pgfutil@tempdima}{0\pgfutil@tempdima}}%
    \pgfpathmoveto{\pgfqpoint{-6pt}{-6pt}}%
    \pgfpathlineto{\pgfqpoint{-6pt}{-6pt}}%
    \pgfpathlineto{\pgfqpoint{-6pt}{6pt}}%
    \pgfpathmoveto{\pgfqpoint{0\pgfutil@tempdima}{0\pgfutil@tempdima}}%
    \pgfpathmoveto{\pgfqpoint{-8pt}{-6pt}}%
    \pgfpathlineto{\pgfqpoint{-8pt}{-6pt}}%
    \pgfpathlineto{\pgfqpoint{-8pt}{6pt}}%
    \pgfusepathqstroke%
}

\pgfarrowsdeclare{zeroone}{zeroone}
{
    \pgfarrowsleftextend{+-.5\pgflinewidth}%
    \pgfarrowsrightextend{+.5\pgflinewidth}%
}
{
    \pgfutil@tempdima=0.6pt%
    \pgfsetdash{}{+0pt}%
    \pgfsetmiterjoin%
    \pgfpathcircle{\pgfpoint{-11.5pt}{0}} {3.5pt}
    \pgfpathmoveto{\pgfqpoint{-6pt}{-6pt}}%
    \pgfpathlineto{\pgfqpoint{-6pt}{6pt}}%
    \pgfusepathqstroke%
}

\makeatother

\tikzset{%
    zig zag to/.style={to path={(\tikztostart) -| ($(\tikztostart)!#1!(\tikztotarget)$) |- (\tikztotarget)}},
    zig zag to/.default=0.5,
    one to zeroone/.style={
        one-zeroone, zig zag to,
    },
    one to one/.style={one-one, zig zag to},
    one to many/.style={one-crow's foot, zig zag to},
    one to omany/.style={one-omany, zig zag to},
    many to one/.style={crow's foot-one, zig zag to},
    many to many/.style={crow's foot-crow's foot, zig zag to},
    blacktable/.style={rectangle split, rectangle split parts=2, draw, align=center},
    redtable/.style={blacktable, fill=gray!20},
    font=\sffamily\small
}

\begin{document}


\title{Multiversion Hindsight Logging for Continuous Training}

\author{Rolando Garcia}
\affiliation{%
      \institution{UC Berkeley}
}
\email{rogarcia@berkeley.edu}

\author{Anusha Dandamudi}
\affiliation{%
      \institution{UC Berkeley}
}
\email{adandamudi@berkeley.edu}

\author{Gabriel Matute}
\affiliation{%
      \institution{UC Berkeley}
}
\email{gmatute@berkeley.edu}

\author{Lehan Wan}
\affiliation{%
      \institution{UC Berkeley}
}
\email{wan_lehan@berkeley.edu}

\author{Joseph Gonzalez}
\affiliation{%
      \institution{UC Berkeley}
}
\email{jegonzal@berkeley.edu}

\author{Joseph M. Hellerstein}
\affiliation{%
      \institution{UC Berkeley}
}
\email{hellerstein@berkeley.edu}

\author{Koushik Sen}
\affiliation{%
      \institution{UC Berkeley}
}
\email{ksen@berkeley.edu}

\begin{abstract}
      Production Machine Learning is a data-intensive process, involving continuous retraining of multiple versions of models, many of which may be running in production at once. When model performance does not meet expectations, Machine Learning Engineers must explore and analyze numerous versions of code, logs and training data to identify root causes and mitigate problems. Traditional software engineering tools fall short in
      this data-rich context.
      FlorDB introduces \emph{multiversion hindsight logging}, a form of acquisitional query processing that allows engineers to use the most recent version's logging statements to query past versions' logs, even when older versions logged different data.
      FlorDB provides a \emph{replay query} interface with accurate cost estimates to help the end-user refine their queries. Once a replay query plan is confirmed, logging statements are propagated across code versions, and the modified training scripts are replayed based on checkpoints from previous runs. Finally,
      FlorDB presents a unified relational view of log history across versions, making it easy to explore behavior across past code iterations. We present a performance evaluation on diverse benchmarks confirming scalability and the ability to deliver real-time query responses, leveraging query-based filtering and checkpoint-based parallelism for efficient replay.
\end{abstract}
\maketitle



\section{Introduction}
Model development for machine learning (ML) is an iterative, experimental and data-intensive process that differs from traditional software engineering.
The primary goal in model development is to boost predictive performance, leading developers to adopt empirical methods involving thorough logging and continuous updates to code and training data~\cite{xin2021production}. 
A recent study highlighted the importance of high-speed experimentation in this field:
as one participant stated, ``the most important thing to do is achieve scary high experimentation velocity''~\cite{shankar2022operationalizing}. 
This approach, however, generates a significant data management challenge for engineers, who must handle numerous iterations of code, datasets, and logs.

For this aggressively agile approach to be successful, machine learning engineers (MLEs) must be able to retroactively examine rich and diverse data from past runs to guide ongoing experiments~\cite{garcia2018context}. 
One method to enable model developers to ``query the past'' is by generating and maintaining comprehensive experiment logs, 
but
this is burdensome and unrealistic. Ensuring comprehensive logs requires (a) foresight to add logging statements to the code for all relevant metrics, without knowing which ones are relevant in advance, and (b) data management discipline for the many log files, code versions, training data and requisite metadata that ensues.

In this paper, we explore a different approach: we show it is possible 
to query the past of machine learning experiments efficiently and on-demand, 
even in the absence of experiment logs. We achieve this automatically on the developer's behalf, by managing model training checkpoints and the corresponding versions of code and data, and exposing experimental history---both what was actually logged \emph{and what could have been logged}---as a virtual database that can be queried with familiar APIs like SQL or dataframe libraries.

\subsection{Multiversion Hindsight Logging}
\label{sec:introhl}

Hindsight logging \cite{garcia2020hindsight} is a lightweight record-replay technique that allows MLEs to add logging statements to a long-running Python program after it executes, and then incrementally \emph{replay} parts of the program very quickly to generate the log outputs that would have emerged had the statements been in the code originally. This technique encourages MLEs to operate optimistically and flexibly with a minimal initial logging scheme, typically restricted to loss and accuracy metrics during the training phase. Additional logging is added \emph{retroactively}, when evidence of issues arises (e.g., in deployment) and further key information is required for debugging.

We realized the concept of hindsight logging in Flor, a record-replay system specifically designed for model training~\cite{garcia2020hindsight}.
Flor embodies two salient features: low-overhead checkpointing, and low-latency replay from checkpoint. Flor checkpointing is adaptive and runs in the background, facilitating low-overhead checkpointing, and limiting the computational resources expended during model training. Simultaneously, the system ensures reduced latency during replay by leveraging memoization and parallelism through checkpoint-resume.

While hindsight logging is a useful core technology, it is insufficient on its own to address typical ML workflows, which invariably involve many training runs. MLEs, in their pursuit of high-velocity experimentation, often wish to revisit not just the most recent run of an experiment, but also prior runs that were performed using different versions of code and data. This presents a complex challenge that goes beyond the capabilities of traditional logging and debugging tools.

To tackle this challenge, in this paper we introduce \emph{multiversion hindsight logging}, and a system called \emph{FlorDB} that provides an efficient and easy-to-use solution. Multiversion hindsight logging is designed to track and manage multiple versions of ML experiments. In doing so, it provides a more comprehensive and effective solution than simple hindsight logging, ensuring that MLEs can readily look back through their iterative development processes, thereby better understanding and learning from their experimentation history. Achieving these goals requires overcoming a set of technical challenges described below.

\subsection{Contributions}
FlorDB, a Multiversion Hindsight Logging system, provides a tabular query interface, treating every run of a version as a set of rows, and 
each logging statement as a ``virtual column''. 
FlorDB includes automatic version control through Git and is designed to interoperate with multiple ML experiment management tools. In addressing the inherent challenges of implementing Multiversion Hindsight Logging, FlorDB makes the following key contributions:

\begin{enumerate}
    \item \textbf{Historical Query Manager using a Unified Relational Model}: FlorDB presents MLEs with a simple metaphor of a relational view of queryable log results—whether the log statements were previously materialized, or need to be generated on demand via hindsight logging. The unified relational model allows users to issue ad-hoc queries using familiar SQL or pandas, thereby simplifying the process of exploring historical code executions . 
    
    \item \textbf{Multiversion Hindsight Logging as Acquisitional Query Processing}: 
    FlorDB extends the concept of Acquisitional Query Processing (AQP)~\cite{madden2005tinydb} by seamlessly integrating data acquisition into query execution through experiment replay. This framework differs from traditional systems that query static datasets. Selected experimental versions undergo sequential processing: logging statements are propagated to generate required data, followed by the partial or parallel replay of modified training scripts.
    
    \item \textbf{Accurate Time Estimation for Query Refinement}: 
    FlorDB incorporates highly accurate cost prediction for estimating the runtime of on-demand replay queries. By leveraging runtime profiling statistics collected during the initial record phase, FlorDB provides highly accurate (error $\leq 5\%$) replay cost estimates. This precision enables developers to refine their queries \emph{before} being bogged down by unnecessarily lengthy execution times.
\end{enumerate}

FlorDB's integrated features provide machine learning engineers with a flexible relational abstraction to capture and query the extended histories of their ML experiments. 
By harnessing the framework of FlorDB, MLEs can more quickly iterate from prior attempts to successful models.

\section{Scenario: Catastrophic Forgetting}\label{sec:scenario}

\begin{figure*}[t!]
      \includegraphics[width=0.98\textwidth]{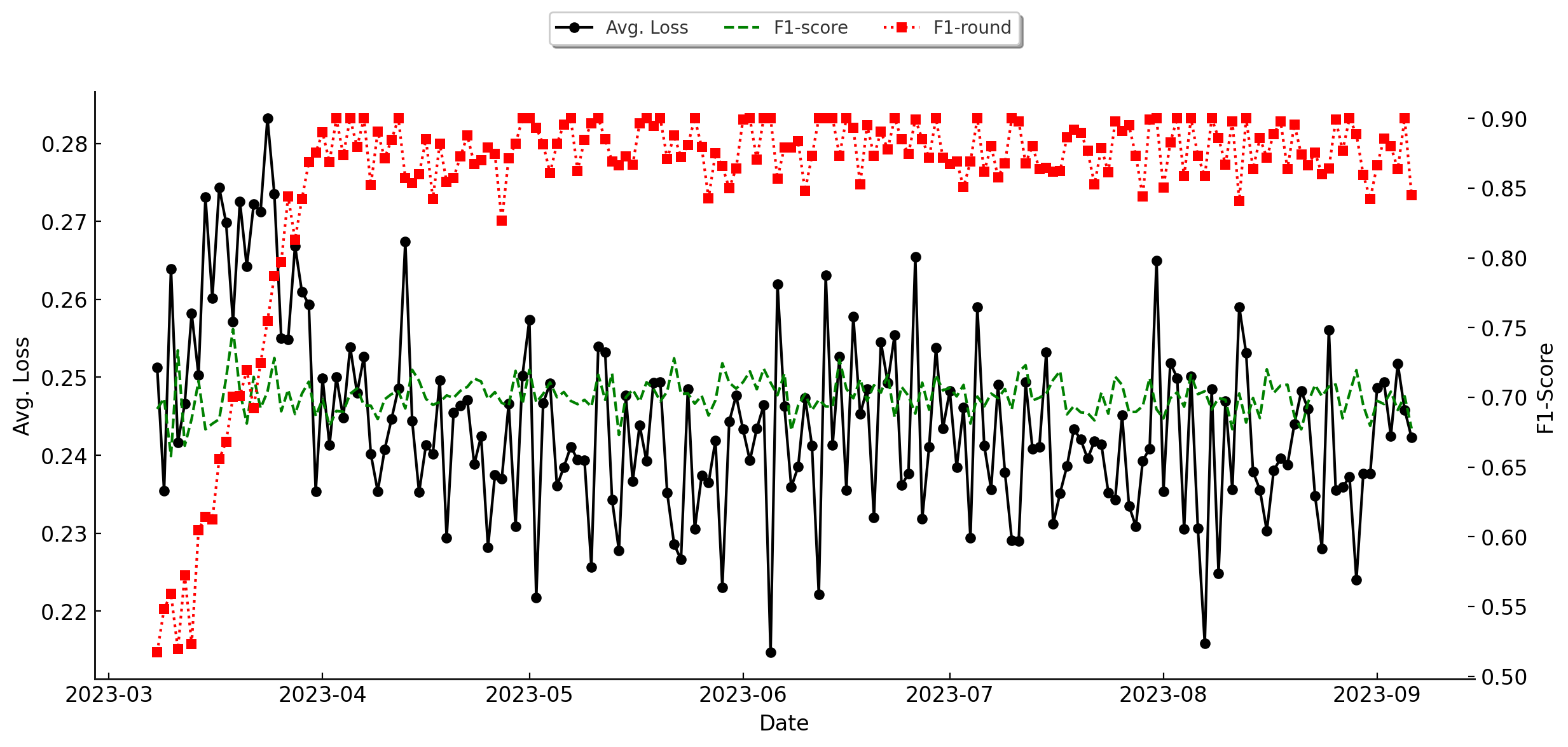}
      \caption{Average training losses and F1-scores for Alice's object detection model over the last 6 months. The model undergoes continuous training, with batches of labeled data added approximately twice a month. These batch dumps result in temporary fluctuations in the loss. \texttt{F1-round} is the F1-score for roundabouts; \texttt{F1-score} is the global F1-score.}
      \label{fig:lossy}
\end{figure*}

\begin{figure*}[h!]
    \centering
    \begin{subfigure}[b]{0.24\textwidth}
        \includegraphics[width=\textwidth]{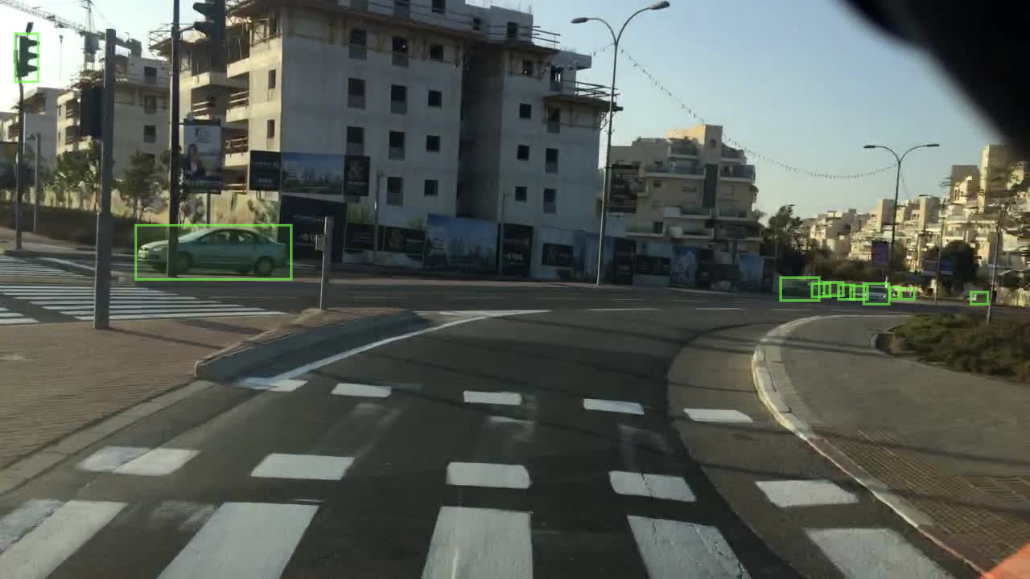}
    \end{subfigure}
    \hfill \vspace{5pt}
    \begin{subfigure}[b]{0.24\textwidth}
        \includegraphics[width=\textwidth]{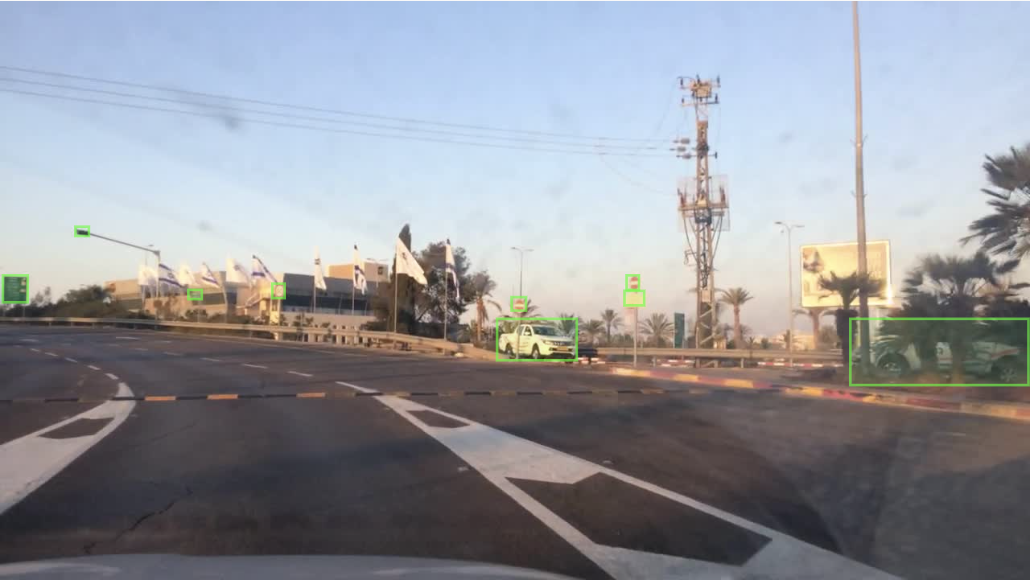}
    \end{subfigure}
    \hfill
    \begin{subfigure}[b]{0.24\textwidth}
        \includegraphics[width=\textwidth]{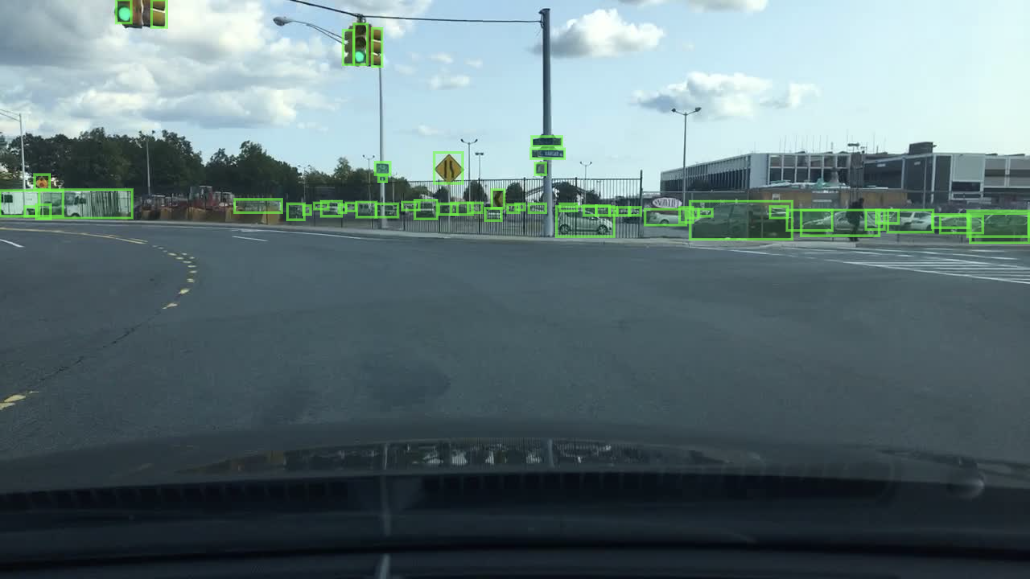}
    \end{subfigure}
    \hfill
    \begin{subfigure}[b]{0.24\textwidth}
        \includegraphics[width=\textwidth]{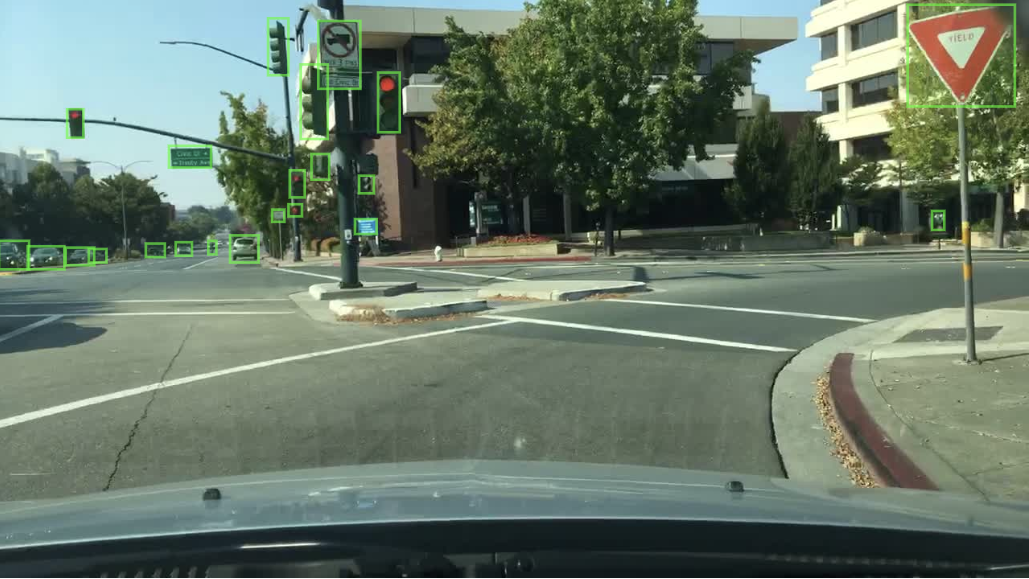}
    \end{subfigure}
    \vspace{10pt} 
    \begin{subfigure}[b]{0.24\textwidth}
        \includegraphics[width=\textwidth]{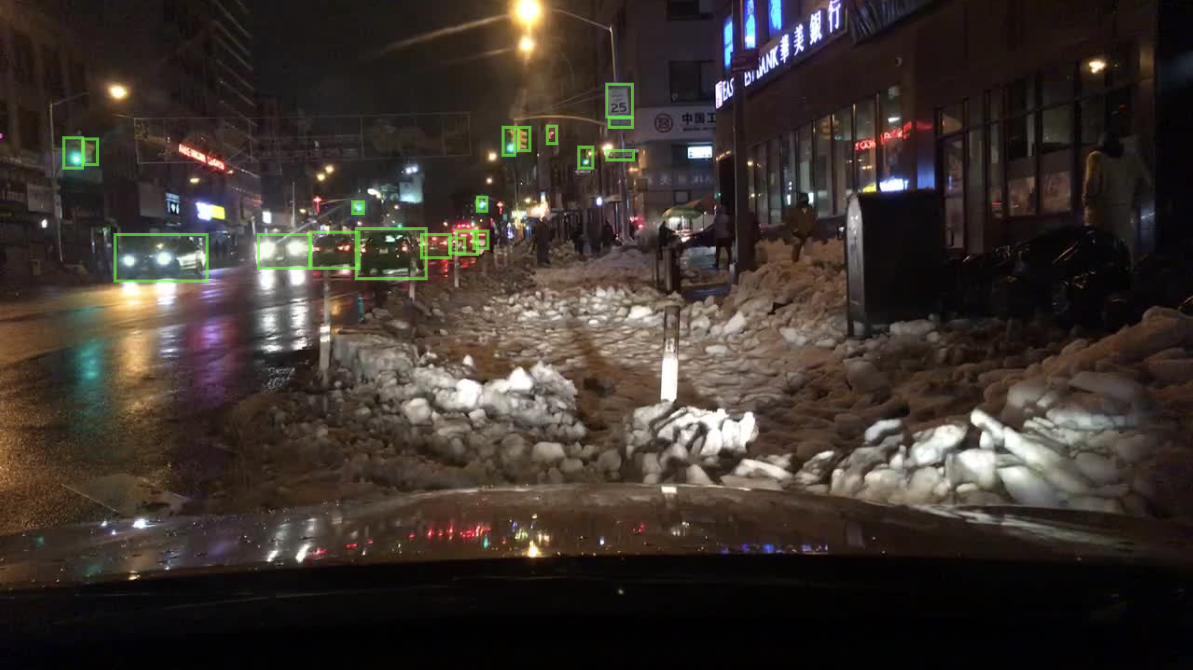}
    \end{subfigure}
    \hfill
    \begin{subfigure}[b]{0.24\textwidth}
        \includegraphics[width=\textwidth]{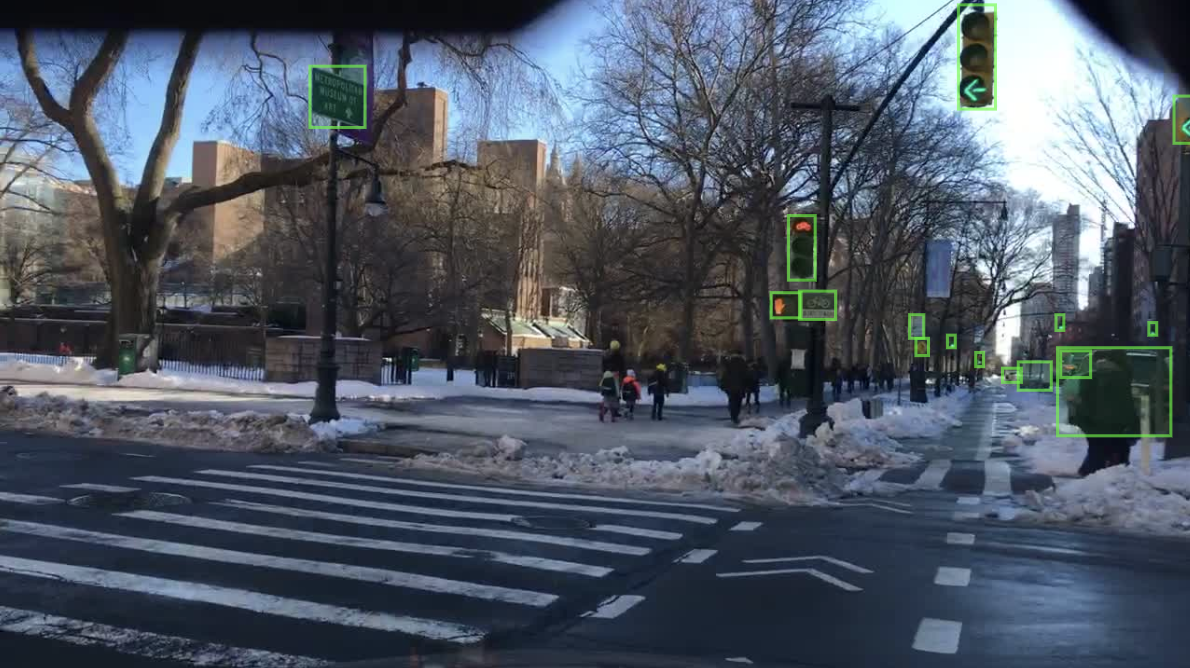}
    \end{subfigure}
    \hfill
    \begin{subfigure}[b]{0.24\textwidth}
        \includegraphics[width=\textwidth]{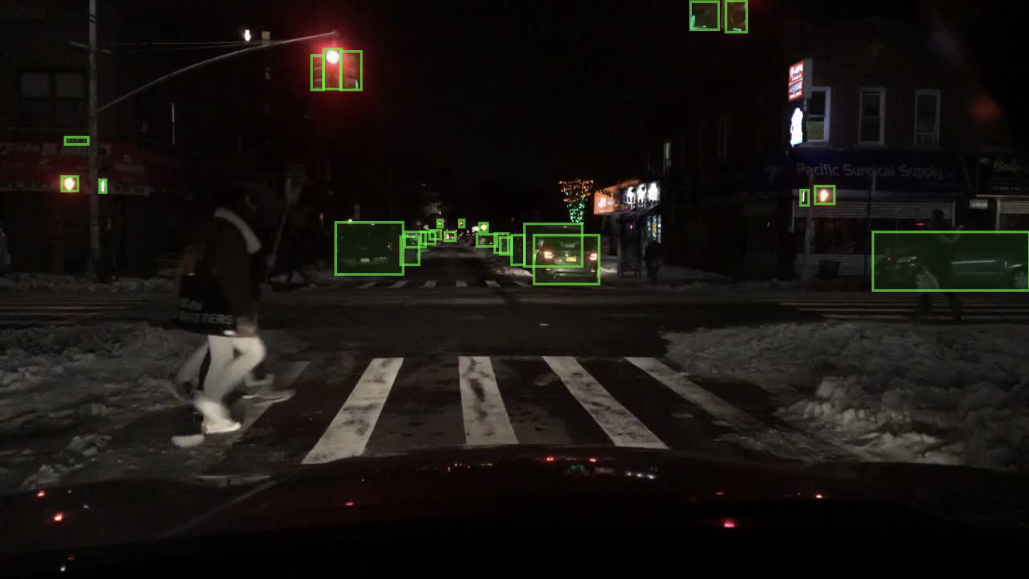}
    \end{subfigure}
    \hfill
    \begin{subfigure}[b]{0.24\textwidth}
        \includegraphics[width=\textwidth]{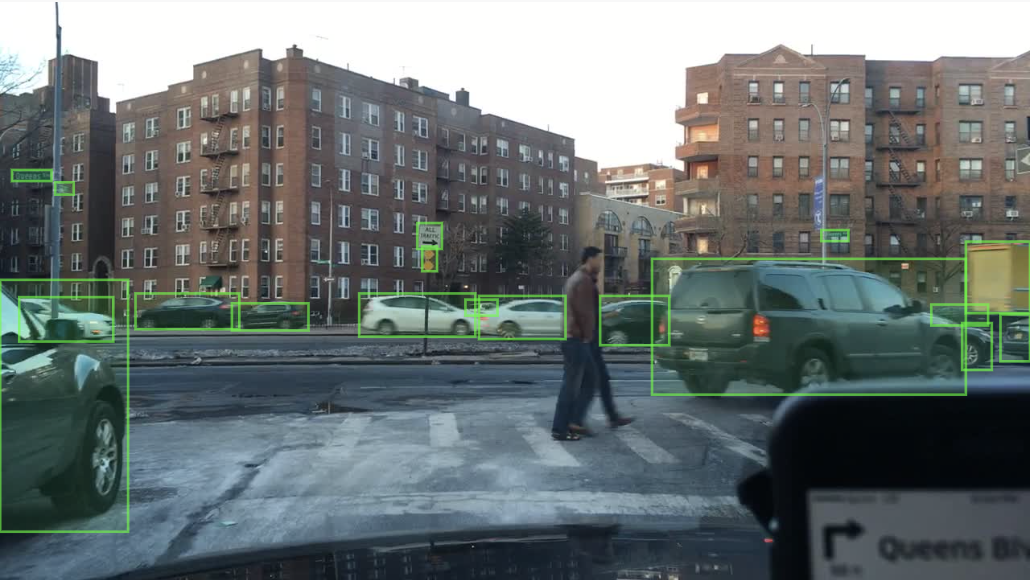}
    \end{subfigure}
    \caption{BDD100K dashcam images with bounding boxes. Top row contains sample images used by Alice for fine-tuning on roundabouts; bottom row corresponds to images for which the model fails to detect pedestrians.}
    \label{fig:2x4grid}
\end{figure*}
For readers who are not MLEs, we motivate FlorDB with an example of
\emph{catastrophic forgetting}, a pressing ML issue that is especially problematic in production settings~\cite{shankar2022operationalizing}. This phenomenon occurs when a model, upon being fine-tuned for a new task, loses its ability to perform well on a previously learned task. This creates a balancing act between updating the model (for new data or tasks) and maintaining its performance (for older tasks). 





\subsection{Object Detection in Autonomous Vehicles}

Imagine an MLE named Alice, responsible for the ongoing fine-tuning of an object detection model used in autonomous vehicles. Operating in a dynamic environment, Alice frequently updates the model with bi-weekly batches of labeled data. Her current focus is the specialized task of object detection within traffic roundabouts. Standard metrics like loss functions and F1-scores, shown in Figure~\ref{fig:lossy}, serve as her evaluation benchmarks.

However, Alice encounters an unexpected problem. A colleague, Bob, informs her of persistent failures in pedestrian detection that have arisen in her new model versions. Upon examining a representative set of images (refer to the bottom row of Figure~\ref{fig:2x4grid}), she verifies the issue.

\subsection{Responding to Performance Regressions}
To combat this specific case of catastrophic forgetting in pedestrian detection, Alice employs hindsight logging for two key actions:

\begin{itemize}
    \item \textbf{Model Rollback}: Alice would like to add hindsight logging statements to calculate the F1 score \textit{for pedestrian detection} across previous model versions. This would allow her to identify the latest point where the model still successfully detected pedestrians, enabling an informed rollback.
    \item \textbf{Alert Tuning}: To prevent the issue from recurring, Alice 
    wants to add alerting logic that checks the value of the pedestrian detector F1 score. However she needs to avoid ``alert fatigue'', when too many spurious alerts are issued. To ensure that her altering is calibrated, she again utilizes hindsight logging to ensure the F1 thresholds for alerting is not triggered by valid runs. 
\end{itemize}

Alice's scenario underscores a significant limitation in current approaches to model versioning and logging. Given the expansive history of model versions, it is neither practical nor efficient for Alice to manually revisit past iterations of training to insert additional logging statements, replay training, and analyze the results. Multiversion hindsight logging provides the high performance replay mechanisms for retroactive logging statement replay with automated back-propagation of those statements to previous versions of code for a comprehensive evaluation. This enables Alice to retroactively analyze metrics (e.g. F1 scores specific to pedestrian detection) across all historical versions.
\lstset{style=custompython}


% ...


\begin{figure}
    \centering
    \includegraphics[width=\columnwidth]{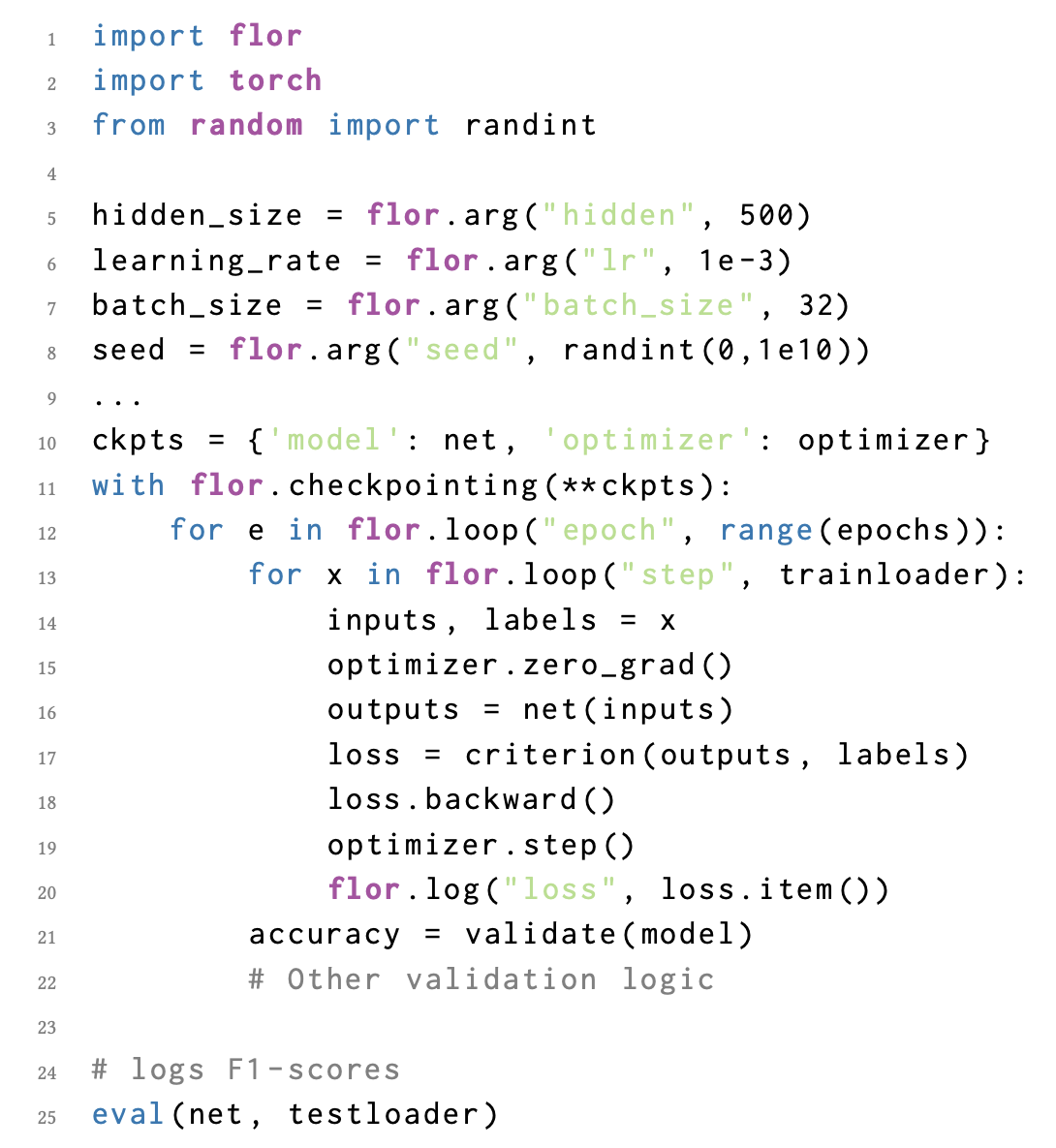}
    \caption{Alice's PyTorch training with Flor API.}
    \label{fig:florhowto}
\end{figure}

\begin{figure*}[t!]
    \centering
    \begin{minipage}[c]{0.8\textwidth}
        \centering
        \resizebox{\textwidth}{!}{%
        \begin{tabular}{ccccccccc||c}
            \toprule
projid & tstamp & filename & hidden & lr & batch\_size & seed & f1\_score & f1\_round & \emph{\textcolor{brewergreen}{f1\_ped}} \\
\midrule
roundabouts & 2023-06-23 & train.py & 500 & 0.001 & 32 & 81 & 0.7022215 & 0.90000000 & \textcolor{brewergreen}{0.5192609} \\
roundabouts & 2023-06-24 & train.py & 500 & 0.001 & 32 & 63 & 0.7043473 & 0.88589064 & \textcolor{brewergreen}{0.5590010} \\
roundabouts & 2023-06-29 & train.py & 500 & 0.001 & 32 & 157 & 0.6912616 & 0.86693724 & \textcolor{brewergreen}{0.5482590} \\
roundabouts & 2023-06-30 & train.py & 500 & 0.001 & 32 & 42 & 0.6994197 & 0.89759576 & \textcolor{brewergreen}{0.5170792} \\
\dots & \dots & \dots & \dots & \dots  & \dots & \dots & \dots  & \textcolor{brewergreen}{\dots} \\
roundabouts & 2023-08-27 & train.py & 500 & 0.001 & 32 & 213 & 0.6982518 & 0.84088466 & \textcolor{brewergreen}{0.2392322} \\
roundabouts & 2023-08-28 & train.py & 500 & 0.001 & 32 & 12 & 0.6770945 & 0.90000000 & \textcolor{brewergreen}{0.2493409} \\
roundabouts & 2023-08-29 & train.py & 500 & 0.001 & 32 & 333 & 0.7026935 & 0.86367500 & \textcolor{brewergreen}{0.2493730} \\
roundabouts & 2023-08-30 & train.py & 500 & 0.001 & 32 & 99 & 0.6818689 & 0.88571969 & \textcolor{brewergreen}{0.2351695} \\
roundabouts & 2023-08-31 & train.py & 500 & 0.001 & 32 & 475 & 0.6995291 & 0.87202768 & \textcolor{brewergreen}{0.2285919} \\
roundabouts & 2023-09-01 & train.py & 500 & 0.001 & 32 & 198 & 0.6850775 & 0.88872464 & \textcolor{brewergreen}{0.2265961} \\
roundabouts & 2023-09-02 & train.py & 500 & 0.001 & 32 & 94 & 0.7200135 & 0.89775851 & \textcolor{brewergreen}{0.2485293} \\
roundabouts & 2023-09-03 & train.py & 500 & 0.001 & 32 & 37 & 0.7030796 & 0.86352228 & \textcolor{brewergreen}{0.2305369} \\
roundabouts & 2023-09-04 & train.py & 500 & 0.001 & 32 & 292 & 0.7084305 & 0.89572726 & \textcolor{brewergreen}{0.2373573} \\
roundabouts & 2023-09-05 & train.py & 500 & 0.001 & 32 & 70 & 0.7090070 & 0.87411934 & \textcolor{brewergreen}{0.2364624} \\
roundabouts & 2023-09-06 & train.py & 500 & 0.001 & 32 & 553 & 0.6865721 & 0.84508553 & \textcolor{brewergreen}{0.2418998} \\
            \bottomrule
        \end{tabular}
        }
    \end{minipage}
    \caption{Alice's \lstinline|flor.dataframe| before and after hindsight logging (black and green text, respectively).}
    \label{fig:experiment_history}
\end{figure*}

\section{User Experience \& API}\label{sec:iterativeBatch}
In this section, we explore how Alice leverages FlorDB\footnote{Available on PyPI, installable via \texttt{pip install flordb}} for addressing performance regressions in pedestrian detection, notably without pre-logged F1-scores for pedestrians. FlorDB enables her to effectively identify and revert to a more efficient model version, and fine-tune alert thresholds. Our discussion underscores the FlorDB API's role in simplifying the ML workflow, facilitating rapid iteration and in-depth analysis for users with varied expertise. We illustrate the API's functionality in key activities such as logging, parameter management, and hindsight logging for model refinement. Key API features include:

\begin{itemize}
    \item \lstinline|flor.log(name: str, value: T) -> T| \\
    Logs a given value with the specified name. Useful for tracking variables or parameters during an experiment run. 
    \item \lstinline|flor.arg(name: str, default: T) -> T| \\
    Reads a value from the command line if provided; else uses the default value. During replay, retrieves logged values from history. Good for setting configuration parameters and allowing for quick adjustments via the command line.
    \item \lstinline|flor.loop(name:str, vals:Iterable[T]) -> Iterable[T]| \\ 
    Functions as a Python generator that maintains global state between iterations. Useful for ``indexing'' replay and coordinating checkpoints.
    \item \lstinline|flor.checkpointing(**kwargs) -> ContextManager[None]| \\ 
    Scopes the set of objects to be checkpointed, such as a model or optimizer, via a Python context manager. Checkpointing is done adaptively at \lstinline|flor.loop| iteration boundaries, based on the checkpoint size and the loop iteration time~\cite{garcia2020hindsight}.
    \item \lstinline|flor.dataframe(*args) -> pd.DataFrame| \\
    Produces a Pandas DataFrame of FlorDB log information with a column corresponding to each argument in \lstinline|*args|. The DataFrame also contains ``dimension'' columns, such as the project id, version id, timestamp, and so on. It is the default view in FlorDB for querying log data and selecting model checkpoints to load, using either Pandas or DataFrame-compatible SQL engines like DuckDB~\cite{raasveldt2019duckdb}.
\end{itemize}

\subsection{Running Experiments}
Returning to our scenario,
Alice executes her model training via the following command that starts her \texttt{train.py} script with specific hyper-parameters:

\begin{verbatim}

  python train.py --kwargs lr=0.001 batch_size=32

\end{verbatim}

Alice's script (\Cref{fig:florhowto}) uses native Python functionality to run her ML experiment. 
Hyper-parameters such as learning rate and batch size are passed in using standard command-line arguments, 
which makes hindsight logging seamless and capable of interoperating with other tools typically used in the ML workflow. 

\subsubsection{Args \& Logs}
In the context of running experiments, Alice utilizes the Flor API to enhance configurability and traceability. The \lstinline|flor.arg| function  (lines 5-8 in \Cref{fig:florhowto}) is instrumental in parameterizing her experiments, allowing her to define hyper-parameters such as \lstinline{hidden_size}, \lstinline{learning_rate}, \lstinline{batch_size}, and \lstinline{seed} either via the command-line or by utilizing default values specified within the script. 
This ensures adaptability and reproducibility of her experiment's configuration.
In addition, \lstinline|flor.log| is employed  (line 20 in \Cref{fig:florhowto}) to record the experiment's loss metrics and associated metadata at each optimizer step. The logging of the loss captures the dynamics of the model's performance, enabling Alice to monitor the training process.

\subsubsection{Checkpointing on Loop Boundaries}
Alice's code defines the set of objects to be checkpointed using
\lstinline|flor.checkpointing| (line~11); within that context,
Alice calls \lstinline|flor.loop| (lines~12-13) to use the checkpoints and control record-replay. After each iteration of the outermost loop, which here corresponds to the completion of an ``epoch,'' FlorDB assesses whether to checkpoint. This decision is based on a balance between the checkpoint's size and the iteration's duration, adhering to a strategy from prior work \cite{garcia2020hindsight}. 
Checkpoints are created adaptively, approximately once per epoch.

\subsubsection{Automatically commit changes to Git} At the end of the experiment, 
FlorDB writes a JSON logfile containing the execution's sequence of log records to the experiment repository, 
and commits all changes. This is so future users of the repository can view historical logs and arguments,
and better reproduce the experiments. 
By committing after every run, Flor ensures historical versions of code
are later available for hindsight logging. 

\subsection{Experiment Analysis and Hindsight Logging}\label{sec:hindsight_logging}

Alice reviews FlorDB's \lstinline|flor.dataframe| output, comparing F1-scores across datasets (see \Cref{fig:experiment_history}).
After receiving insights from her colleague, Alice shifts her analysis focus to the pedestrian F1-scores.
By editing \texttt{train.py} to include a new \lstinline|flor.log| for \texttt{f1\_ped}, 
Alice leverages hindsight logging to populate this metric retroactively. 
She invokes the replay feature via CLI as follows:

\begin{verbatim}

python -m flor replay f1_ped "tstamp >= '2023-06-23'"

\end{verbatim}
This command extends \texttt{f1\_ped} logging to any past experiments dated June 23, 2023 or later, in preparation for subsequent analysis.

Upon initiating the replay command, 
Alice is shown a schedule of experiments slated for replay and must confirm to proceed. 
After receiving confirmation, Flor prepares to replay the selected prior runs: it restores their versions from github, adds the new \texttt{f1\_ped} logging statement to each, and executes all versions using the configuration variable values originally logged via \lstinline|flor.arg|.
Notably, during this replay, Flor's hindsight logging functionality can bypass the main loop (lines 12-22 in \Cref{fig:florhowto}) to directly load the model's final state, 
since the \texttt{f1\_ped} statement is computed post-training (line 25 in \Cref{fig:florhowto}).

Once replay is finished, Alice views the updated results with the following command:
\begin{lstlisting}[style=custompython, xleftmargin=8pt, numbers=none]
cols = ['hidden', 'lr', 'batch_size', 'seed'] 
cols += ['f1_score', 'f1_round', 'f1_ped']
flor.dataframe(*cols)
\end{lstlisting}
This command now generates a table including the \texttt{f1\_ped} metric (green text in \Cref{fig:experiment_history}) for entries dated June 23, 2023, or later. 
This new data reveals a previously unnoticed performance regression in pedestrian classifications, as Bob had indicated (\texttt{f1\_ped} values in the range \texttt{tstamp} $\geq$ \texttt{08-27}).

To further refine the analysis and calibrate alert thresholds, 
Alice updates the validation function to add a new \lstinline|flor.log| statement
as part of an assertion that pedestrian F1-scores never fall below some threshold. 
For retroactive application, she again employs FlorDB's replay feature:
\begin{verbatim}
python -m flor replay f1_alert \
"tstamp >= '2023-06-23' and tstamp < '2023-07-01'"
\end{verbatim}
This process selectively executes each epoch's validation logic (lines 21-22), 
avoiding retraining (lines 13-20) by loading final epoch states from checkpoints.
FlorDB's control via \lstinline|flor.loop| facilitates this selective process: 
the outer loop cycles through epochs using checkpoints, while the inner loop is bypassed.

\begin{figure*}[t!]
    \centering
    \includegraphics[width=0.9\textwidth]{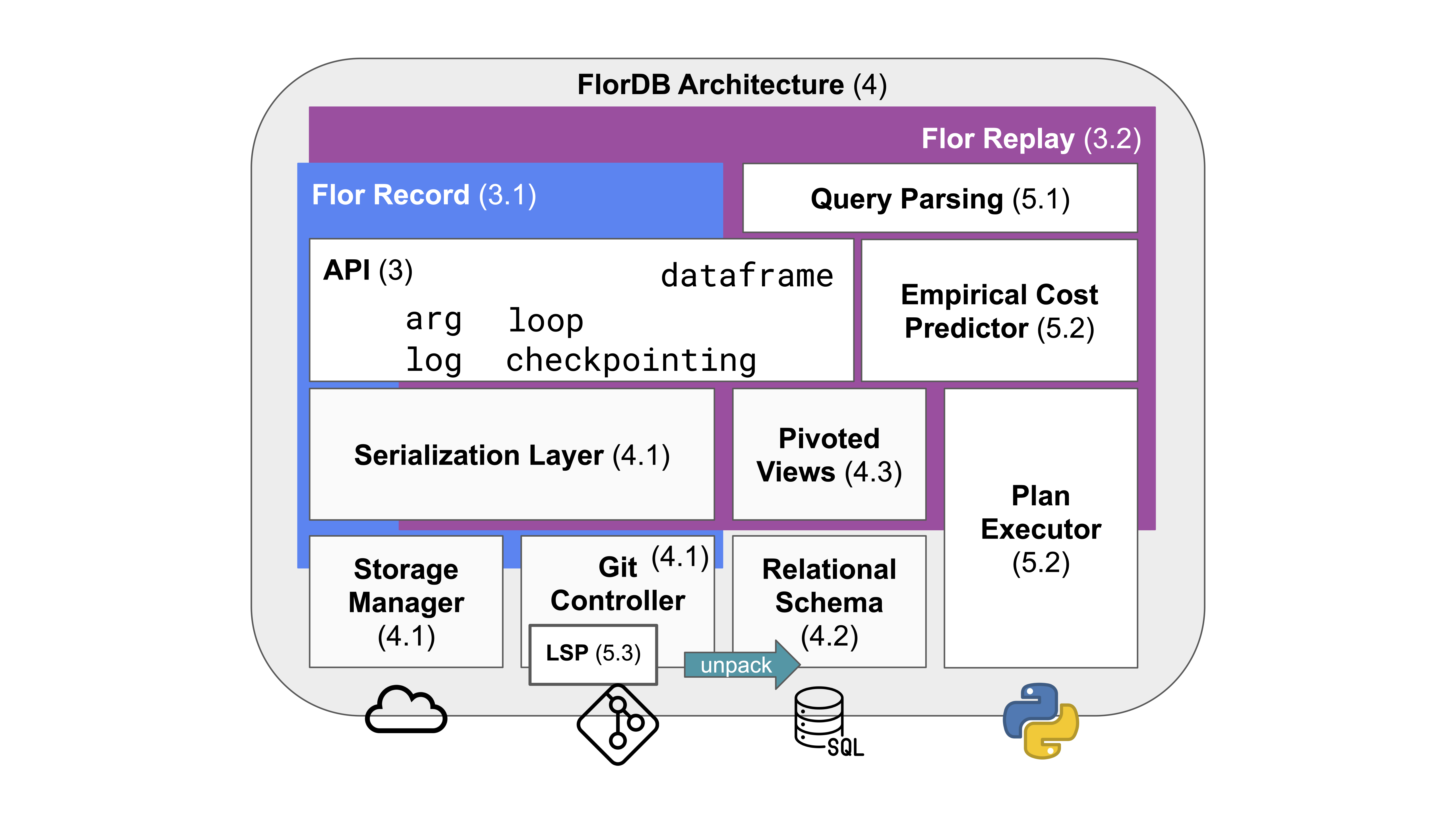}
    \caption{FlorDB architecture diagram with subsection headers in parentheses.}
    \label{fig:archDiagram}
\end{figure*}

\section{System Architecture}\label{sec:sys_arch}
The design of FlorDB aims to provide a unified platform for the management of code, checkpoints, logs, and their evolution over time (\Cref{fig:archDiagram}). 
FlorDB maintains a cross-referenced relational database that interconnects code versions in git, large files in your preferred storage solution (e.g. an S3 bucket), and metadata stored in the database.
This section outlines the mechanisms FlorDB uses for executing experiments and replaying them for hindsight logging, highlighting how it stores, manages and tracks the data and metadata of model training.

\subsection{Storage and Data Layout}

As shown in the bottom of \Cref{fig:archDiagram}, FlorDB's storage architecture consists of the following three units:

\begin{itemize}
	\item \textbf{Git Repository}: For code version control, FlorDB uses git to capture the state of the working directory and JSON log files post-execution. FlorDB and the user commit to the same Git repository,
    but FlorDB auto-commits to a dedicated branch for added safety --- the user may treat that branch as any other branch.
    This facilitates a comprehensive  tracking of code versions and more transparent sharing of execution metadata, enhancing experiment reproducibility and analysis.
	\item \textbf{Relational Database}: Following execution and logging to JSON, FlorDB unpacks and normalizes data from the JSON logs to populate a relational database (which is pluggable; FlorDB uses SQLite by default). This process transforms the semi-structured JSON data into a structured format, suitable for efficient querying and analysis.
	\item \textbf{Object Storage}: This component is responsible for storing large files and checkpoints, integrating with the database system to offer scalable and flexible storage solutions, whether locally or in the cloud.
\end{itemize}

This heterogeneous approach to storage ensures that FlorDB is well-equipped to manage the complexities of code versioning, data checkpointing, and metadata storage.

\subsubsection{Serialization Layer}
The API utilizes a serialization layer to prepare objects for storage. 
This layer differentiates between object types --- PyTorch objects are serialized using native PyTorch functionality, 
while cloudpickle provides a general-purpose serialization solution.
For in-depth information on how fork and copy-on-write are used for background serialization, 
the reader is referred to a technical report~\cite{Liu:EECS-2020-79}.

\subsection{Relational Schema}\label{sec:datamodel}
\begin{figure}[t!]
  \centering
  \resizebox{0.99\columnwidth}{!}{
    \begin{tikzpicture}[node distance=1.5cm and 1.5cm, blacktable/.style={rectangle split, rectangle split parts=2, draw, align=center},
                    bluetable/.style={blacktable, fill=brewergreen},
                    redtable/.style={blacktable, fill=gray!20}, font=\sffamily\small]

                 \node[blacktable] (loops) {loops \nodepart{two}
                        \begin{tabular}{ll}
                              \small \underline{ctx\_id}:            & \small integer     \\[-0.2em]
                              \small parent\_ctx\_id:      & \small integer     \\[-0.2em]
                              \small loop\_name:        & \small text     \\[-0.2em]
                              \small loop\_entries:          & \small integer  \\[-0.2em]
                              \small loop\_iteration:   & \small integer
                        \end{tabular}};

                  \node[blacktable, left=of loops] (logs) {logs \nodepart{two}
                        \begin{tabular}{ll}
                              \small \underline{projid}:   & \small text     \\[-0.2em]
                              \small \underline{tstamp}:   & \small datetime \\[-0.2em]
                              \small \underline{filename}: & \small text     \\[-0.2em]
                              \small \underline{ctx\_id}: & \small integer     \\[-0.2em]
                              \small \underline{value\_name}:     & \small text     \\[-0.2em]
                              \small value:    & \small text     \\[-0.2em]
                              \small value\_type:     & \small integer
                        \end{tabular}
                  };
                        
                  \node[redtable, above=of logs] (ts2vid) {ts2vid \nodepart{two}
                        \begin{tabular}{ll}
                              \small \underline{ts\_start}: & \small datetime \\[-0.2em]
                              \small vid:       & \small text     \\[-0.2em]
                              \small ts\_end:   & \small datetime
                        \end{tabular}};
                  \node[redtable, right=of ts2vid] (repo) {git \nodepart{two}
                        \begin{tabular}{ll}
                              \small \underline{vid}:        & \small text    \\[-0.2em]
                              \small \underline{filename}:   & \small text    \\[-0.2em]
                              \small parent\_vid: & \small text \\[-0.2em]
                              \small contents:   & \small text
                        \end{tabular}};
                  \node[redtable, below=of logs] (obj_store) {obj\_store \nodepart{two}
                        \begin{tabular}{ll}
                              \small \underline{projid}:   & \small text     \\[-0.2em]
                              \small \underline{tstamp}:   & \small datetime \\[-0.2em]
                              \small \underline{filename}: & \small text     \\[-0.2em]
                              \small \underline{ctx\_id}: & \small integer     \\[-0.2em]
                              \small \underline{value\_name}:     & \small text     \\[-0.2em]
                              \small contents:    & \small blob     \\[-0.2em]
                        \end{tabular}};

                  \draw[one to omany] 
                  ($(logs.east)+(0,0pt)$) --
                  ($(loops.west)+(0,0pt)$);
  
                \draw[one to zeroone] 
                ($(loops.north)+(-25pt,0)$) 
                to [out=160,in=20,looseness=3] 
                ($(loops.north)+(25pt,0)$);

                  \draw[one to zeroone] ($(logs.south)-(0,0pt)$) -- ($(obj_store.north)-(0,0pt)$);
                  \draw[one to many]
                  ($(logs.north)-(0,0pt)$) --($(ts2vid.south)+(0,0pt)$);
                  
                  \draw[one to many] 
                  ($(ts2vid.east)-(0,0pt)$) -- ($(repo.west)+(0,0pt)$);
                  
                  \draw[one to zeroone] 
                ($(repo.north)+(-20pt,0)$) 
                to [out=160,in=20,looseness=3] 
                ($(repo.north)+(20pt,0)$);

            \end{tikzpicture}
  }

  \caption{Data Model Diagram in Crow's Foot Notation. Tables with a white background are basic; those in gray are virtual.}
  \label{fig:datamodel}
\end{figure}
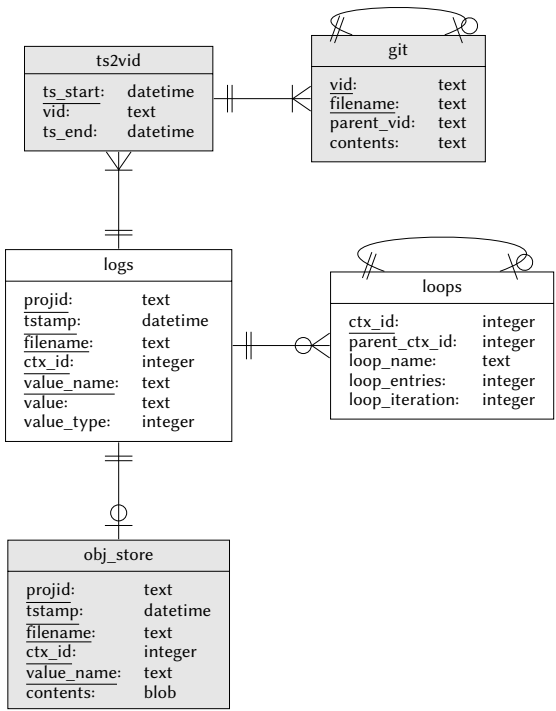

The relational schema of FlorDB is a structured and normalized view over the JSON logs. As depicted in \Cref{fig:datamodel}, the \texttt{logs} table is central to this schema, managing the data and relevant metadata of experiments. The schema for the \texttt{logs} table includes:

\begin{itemize}
    \item \textbf{projid}: Basename of the root-level working directory.
    \item \textbf{tstamp}: Datetime marking when the run was started.
    \item \textbf{filename}: Name of the source file producing the log entry.
    \item \textbf{ctx\_id}: An integer acting as a foreign key to associate each log entry with a related entry in the \texttt{loops} table.
    \item \textbf{value\_name}: text descriptor for the variable being logged.
    \item \textbf{value}: The actual data logged, stored as text.
    \item \textbf{value\_type}: Integer classifying the \emph{type} of data logged.
\end{itemize}

Complementing the \texttt{logs} table, the \texttt{loops} table tracks the context of the execution flow (the call stack) as follows:

\begin{itemize}
    \item \textbf{ctx\_id}: Unique identifier for a specific loop context.
    \item \textbf{parent\_ctx\_id}: References the parent context in nested loops.
    \item \textbf{loop\_name}: The name or identifier of the loop.
    \item \textbf{loop\_entries}: Number of entries or iterations within the loop.
    \item \textbf{loop\_iteration}: Specific iteration count within a loop.
\end{itemize}

The ``unpack'' process involves converting logging objects from JSON format into the relational structure of the database. Each JSON log record is parsed. 
The data is then mapped to the relevant columns in both the \texttt{logs} and \texttt{loops} tables. In the case of a \emph{null} loop indicating a top-level log entry, the log record is mapped solely within the \texttt{logs} table; in cases of nested loops, the JSON object encapsulates the hierarchical structure of an experiment's iterations, and the corresponding log record is reflected in both \texttt{logs} and \texttt{loops} tables.

\subsection{Pivoted Views}\label{sec:dataframe}
The \lstinline[style=custompython]{flor.dataframe} presents the content
of the database in a single-table view (\Cref{fig:pivotViewDef}). 
This view is generated from the relational schema (\Cref{fig:datamodel}) by
joining the \texttt{logs} table with additional dimensions of data, such as the \texttt{loops} table, which are essential for capturing the full context of each experimental run. This is followed by a series of transformations that pivot the combined data, turning relational values (e.g., \texttt{logs.value\_name}, \texttt{loops.loop\_name})
into column headers, to create a wide-format table where each unique loop and log entry name becomes a distinct column in the DataFrame (e.g. \Cref{fig:experiment_history}).

\subsubsection{Domain Mapping}\label{sec:domainmapping}
We use the \lstinline[style=custompython]{flor.dataframe} as the default
Flor view because of its easy-to-understand application-level semantics.
Specifically:

\begin{itemize}
    \item \textbf{Running Experiments $\rightarrow$ Adding Rows:} The execution of a new experiment results in the addition of rows to \lstinline[style=custompython]{flor.dataframe}. Each row in the DataFrame signifies a discrete iteration within an experiment, collating related data such as metrics, hyper-parameters, and state.
    \item \textbf{Adding Logging Statements $\rightarrow$ Adding Virtual Columns:} The inclusion of logging statements in the source code induces FlorDB to add virtual columns to the view produced by \lstinline[style=custompython]{flor.dataframe}. 
    These virtual columns can then be referenced in subsequent queries like any other column in \lstinline[style=custompython]{flor.dataframe}.
    \item \textbf{Replay from Checkpoint $\rightarrow$ Backfilling Nulls:} Replaying an experiment from checkpoints corresponds to the backfilling of \emph{null} values in the  \lstinline[style=custompython]{flor.dataframe} view. This capability is critical to achieving the hindsight logging abstraction that FlorDB supports.
\end{itemize}

The ML developer can envision the \lstinline[style=custompython]{flor.dataframe} as a universal relational view over the columns they have selected, 
where the rows represent individual experiment runs or, in the presence of loops, iterations within those runs. The columns in the \lstinline[style=custompython]{flor.dataframe} view are potentially infinite; they can be defined post-hoc and subsequently populated using replay from a checkpoint. This fluidity in defining and back-filling columns and data allows a relational model to adapt to an experimenter's evolving analytical needs.

\begin{figure}[!t]
    \centering
    \begin{lstlisting}[style=custompython, numbers=none, showstringspaces=false]
    def dataframe(conn, *args):
        dataframes = []
        loops = pd.read_sql(
          "SELECT * FROM loops", conn)
        for val_name in args:
            logs = pd.read_sql(f"""
              SELECT * FROM logs 
              WHERE value_name = "{val_name}"
              """, conn)
            logs = logs.rename(
              columns={"value": value_name})
            # Unroll loop context
            while logs["ctx_id"].notna().any():
                # Iterate until fixpoint
                logs = pd.merge(
                  left=loops, right=logs,
                  how="inner", on=["ctx_id"])
                ln = logs["loop_name"].unique()
                logs = logs.drop(
                  columns=["loop_name", 
                  "loop_entries"])
                logs = logs.rename(
                  columns={"loop_iteration": ln[0]})
                logs["ctx_id"] = logs["p_ctx_id"]
                logs = logs.drop(
                  columns=["p_ctx_id"])
            logs = logs.drop(columns=["ctx_id"])
            dataframes.append(logs)
        all_joined = reduce(
          outer_join_on_common_columns, dataframes)
        cols = [c for c in all_joined.columns 
          if c not in args] + list(args)
        return all_joined[cols]
\end{lstlisting}
    \caption{Python code implementing data reshaping for \lstinline[style=custompython]{flor.dataframe}. This figure illustrates the sequence of operations—merging, transforming, and pivoting—used to construct the experiment view within FlorDB.}
    \label{fig:pivotViewDef}
\end{figure}

\section{Acquisitional Query Processing}\label{sec:aqp}

FlorDB extends the concept of Acquisitional Query Processing (AQP)~\cite{madden2005tinydb}, which involves efficient data acquisition during query execution. 
AQP was originally presented in the context of live sensing;
here we acquire data by re-running training code from checkpoints.
The process begins at the query construction stage, where criteria for selecting experimental versions are established (\Cref{sec:queryconst}). 
Each selected version has logging statements propagated to it in order to generate the required data (\Cref{sec:statement_prop}). Subsequently, a modified training script is executed with a \lstinline{--replay_flor} flag, enabling replay query operators to perform partial or parallel model training replay (\Cref{sec:Planexecutor}). 
In this section, we describe multiversion hindsight logging via the sequence of mechanisms by which 
FlorDB integrates data acquisition into query execution. This also paves the way for comprehensive post-hoc analysis and continuous model refinement.

\subsection{Query Parsing}\label{sec:queryconst}
The \texttt{flor replay} operation begins with query construction, a step that determines the scope and precision of the retrospective analysis. This stage is crucial for identifying the code versions to be replayed and for setting the level of detail in the logging process.
The query is formulated using the following command structure:
\begin{verbatim}
 python -m flor replay [-h] VARS [where_clause]
\end{verbatim}

The components of the query include:
\begin{itemize}
    \item \textbf{VARS}: This argument specifies the logging variables that the user wishes to generate during the replay. These variables correspond to the log statements that will be regenerated across different code versions.
    \item \textbf{where\_clause}: This optional argument acts as a filter to refine the replay's scope. It can be used to constrain the replay to certain conditions, such as experiments conducted within a specific timeframe or parameter set. The \texttt{where\_clause} is evaluated against a \lstinline[style=custompython]{flor.dataframe} prior to replay. 
\end{itemize}

By judiciously crafting the query, users can ensure that the replay operation is both focused on the necessary aspects of the experiment and efficient in terms of computational resource usage.

\subsection{Empirical Cost Prediction \& Plan Execution}\label{sec:costPredictor}

\begin{figure}[t]
      \includegraphics[width=0.95\columnwidth]{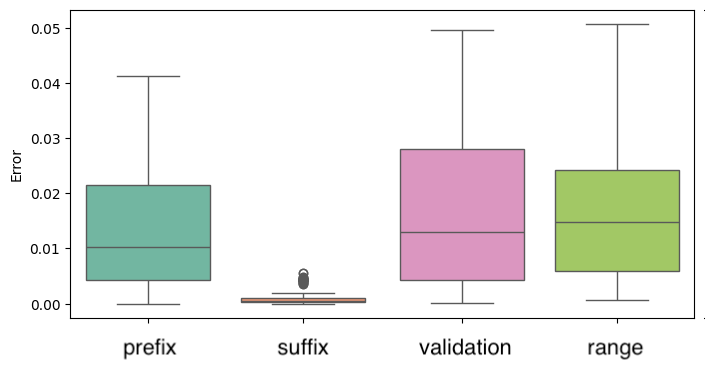}
        \caption{Time estimation error per replay operator; micro-benchmarks correspond to a 2-layer neural network.\protect\footnotemark}
      \label{fig:errBoxPlot}
\end{figure}
\footnotetext{See \texttt{train.py} in \protect\url{https://github.com/ucbepic/ml_tutorial}}

Machine Learning (ML) training scripts can be long-running, and even selective replay can be time-consuming. Therefore, ML engineers need to know how long a replay will take, to decide whether further query refinement is needed. 
Traditional query cost estimation is a notoriously hard problem, often exhibiting orders of magnitude errors in estimation~\cite{leis2015good}. By contrast, we can estimate query runtime for replay executions extremely accurately (\Cref{fig:errBoxPlot}). 
This is because FlorDB profiles runtime information during the record phase.
When initiating a Flor replay, these runtime statistics are queried using the \lstinline[style=custompython]{flor.dataframe} API call. This allows us to reliably predict the time required for replay, enabling users to tailor their queries for efficiency.
We categorize the cost estimation into four levels:

\lstset{
  literate={£}{\$}1
}

\begin{enumerate}
    \item \textbf{Prefix Scan}: Executes statements before the main loop, logging preliminary setup operations (e.g., lines 1-10 in \Cref{fig:florhowto}. This level is useful when examining data preparation and featurization.
    \item \textbf{Suffix Scan}: Executes the initial setup code (as in prefix scan), loads the end state of the outermost loop from a checkpoint, and runs the final script segments (lines 23-EOF in \Cref{fig:florhowto}). This level provides a more detailed view, useful when examining the results of training (e.g. accuracy, recall, F1-scores).
    \item \textbf{Validation Scan}. Steps through the main training loop to execute model validation logic (lines 21-22 in \Cref{fig:florhowto})., loading checkpoints once per epoch but skipping the nested training loop. This level is useful for examining the validation process in detail.
    \item \textbf{Range Scan}: Runs the training loop in depth (lines 12-22 in \Cref{fig:florhowto}) over a selected range to capture nested logging statements --- used for the most granular analyses (e.g. of gradients). A range scan running from epoch 0 to $N$ is considered a full scan. To achieve replay parallelism, FlorDB executes range scans over non-overlapping intervals.
\end{enumerate}

\subsubsection{Plan Execution}\label{sec:Planexecutor}
Once the replay's depth level has been determined by a static analysis of the training script, and the end-user
has accepted the replay plan, a replay subprocess is invoked for each version in the plan, with CLI flags and arguments formatted as strings parameterizing the replay's depth level. 
Flor replay arguments can be used to break out of the execution in a prefix scan upon encountering the first \lstinline[style=custompython]{flor.loop} 
or set to skip the first \lstinline[style=custompython]{flor.loop} loading its end checkpoint as part of a suffix scan; 
or set to step into the the first \lstinline[style=custompython]{flor.loop} as part of a validation scan, 
and so on.
As shown above, validation scans and range scans control access to the outermost loop,
and the nested loop,
a key difference being that in validation scan the nested loop is skipped but not in the range scan.

\lstset{style=custompython}

\begin{table*}
      \centering
      \caption{Computer vision and NLP benchmarks used in our evaluation.}
      \resizebox{.98\textwidth}{!}{%
            \begin{tabular}{|l|l|l|l|l|l|l|}
                  \hline
                  Model                                 & Model Size & Data                                     & Data Size & Objective                & Evaluation      & Application                 \\ \hline \hline
                  ResNet-152~\cite{he2016deep}          & 242 MB     & ImageNet-1k~\cite{imagenet15russakovsky} & 156 GB    & image classification     & accuracy        & computer vision             \\ \hline
                  BERT~\cite{devlinBERT}                & 440 MB     & Wikipedia~\cite{wikidump}                & 40.8 GB   & masked language modeling & accuracy        & natural language processing \\ \hline
                  GPT-2~\cite{radford2019language}      & 548 MB     & Wikipedia~\cite{wikidump}                & 40.8 GB   & text generation          & perplexity      & natural language processing \\ \hline
                  LayoutLMv3~\cite{huang2022layoutlmv3} & 501 MB     & FUNSD~\cite{jaume2019}                   & 36 MB     & form understanding       & F1-score        & document intelligence       \\ \hline
                  DETR~\cite{carion2020endtoend}        & 167 MB     & CPPE-5~\cite{dagli2021cppe5}             & 234 MB    & object detection         & $\mu$-precision & computer vision             \\ \hline
                  TAPAS~\cite{herzig-etal-2020-tapas}   & 443 MB     & WTQ~\cite{pasupat2015compositional}      & 429 MB    & table question answering & accuracy        & document intelligence       \\ \hline
            \end{tabular}
      }
      \label{tab:experiment_list}
\end{table*}

\begin{figure*}[t]
    \centering
    \begin{subfigure}[b]{0.3\textwidth}
        \includegraphics[width=\textwidth]{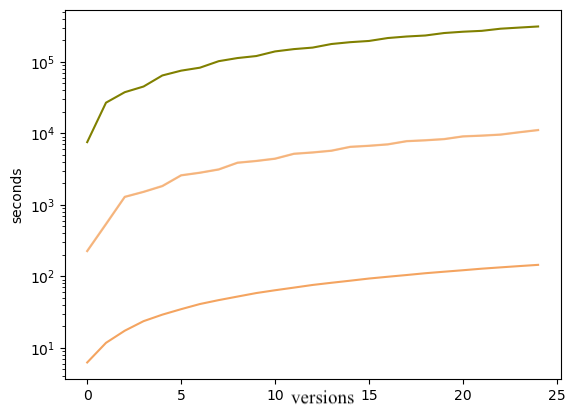}
        \caption{ResNet-152}
    \end{subfigure}\hfill
    \begin{subfigure}[b]{0.3\textwidth}
        \includegraphics[width=\textwidth]{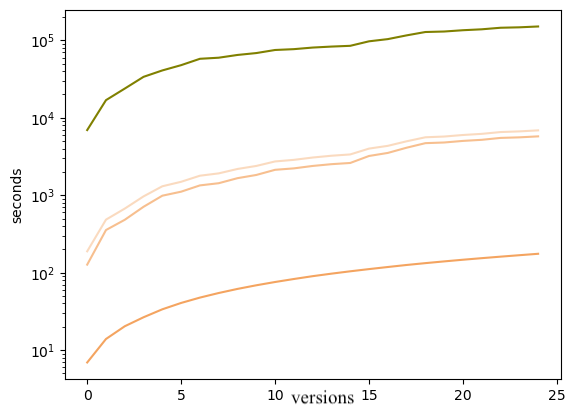}
        \caption{BERT}
    \end{subfigure}\hfill
    \begin{subfigure}[b]{0.3\textwidth}
        \includegraphics[width=\textwidth]{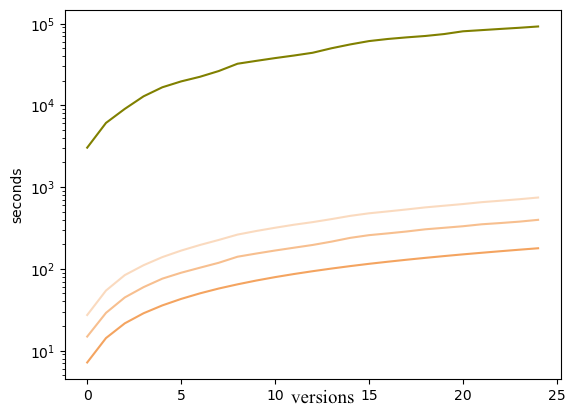}
        \caption{GPT-2}
    \end{subfigure}

    \vspace{1em} 

    \begin{subfigure}[b]{0.3\textwidth}
        \includegraphics[width=\textwidth]{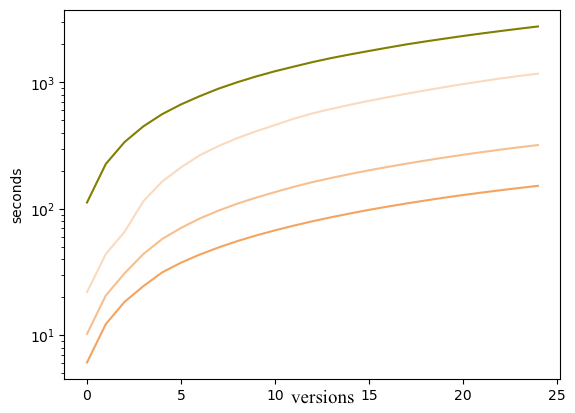}
        \caption{LayoutLMV3}
    \end{subfigure}\hfill
    \begin{subfigure}[b]{0.3\textwidth}
        \includegraphics[width=\textwidth]{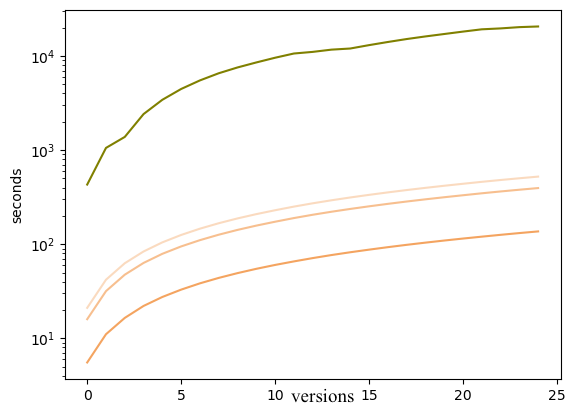}
        \caption{DETR}
    \end{subfigure}\hfill
    \begin{subfigure}[b]{0.3\textwidth}
        \includegraphics[width=\textwidth]{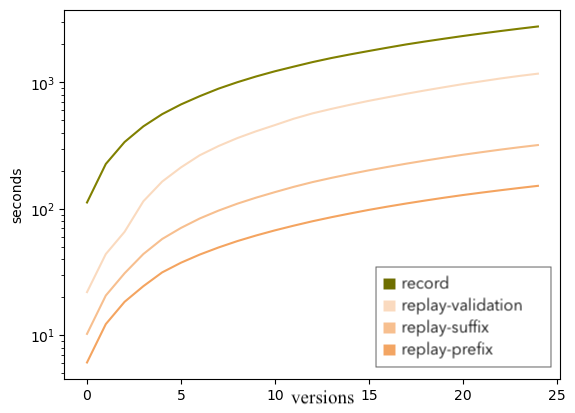}
        \caption{Tapas}
    \end{subfigure}
    \caption{Comparative visualization of processing times for different Flor record-replay operations against the number of experiment versions. Note the logarithmic scale on the y-axis, emphasizing the time disparities between different operations.}
    \label{fig:linePlots}
\end{figure*}

\subsection{Logging Statement Propagation}\label{sec:statement_prop}
After selecting the versions and log variables for replay, 
but before the execution of individual Flor replay subprocesses,
FlorDB adapts the code from the git repository to incorporate new log statements. This involves propagating log statements from a newer version \( Y \) to an older version \( X \) (\( X < Y \)).
The core challenge here is one of code-block alignment: we must identify the most appropriate line
in version \( X \) to insert a logging statement that was originally in line \( L \)
of version \( Y \).
Code alignment is a fundamental task in software engineering,
essential for understanding and managing changes in source files~\cite{horwitz1990identifying, ladd1994semantic}.
It allows developers to track modifications and maintain consistency across different versions
of a file. In this context, GumTree emerges as a state-of-the-art technique for code block alignment.

\subsubsection{Why GumTree is Sufficient for Our Use Case.}
FlorDB's code alignment uses GumTree~\cite{falleri2014fine} for aligning code blocks, aiding in the accurate insertion of log statements across versions. The \lstinline|flor.loop()| function establishes stable anchor points, with main and nested training loops typically serving as consistent markers across versions. This consistency, enhanced by GumTree's alignment capabilities, ensures accurate log statement propagation.
While GumTree is capable of aligning code blocks without Flor API anchors, the presence of these static loop identifiers all but ensures its efficiency and accuracy:
in all instances of logging statement propagation evaluated in \Cref{sec:evaluation}, this combined approach was successful. 


To assess the effectiveness of these techniques, we conducted an evaluation focusing on accuracy and comparing GumTree with the Myers algorithm~\cite{myers1986nd, falleri2014fine}. The evaluation included synthetic benchmarks, randomly mutated PyTorch programs, and ecological benchmarks from GitHub. Our findings, detailed in a technical report \cite{dandamudi_thesis}, show GumTree's superior performance, especially in handling code refactoring and variable renaming.

\section{Evaluation}\label{sec:evaluation}
This evaluation aims to demonstrate the key strengths of FlorDB: its ability to efficiently handle multiversion hindsight logging and its potential for significant speed-ups during replay. We showcase FlorDB's versatility by applying it to a diverse range of model architectures, as detailed in Table~\ref{tab:experiment_list}. This table summarizes each model's characteristics, evaluation metrics, and real-world applications.
Building upon our prior work that established Flor's low-overhead recording and parallel replay capabilities for single versions~\cite{garcia2020hindsight}, this study focuses on FlorDB's ability to generalize and scale effectively across multiple versions. Our evaluation delves into two key areas: scalability and responsiveness. FlorDB's ability to handle growing numbers of model versions is assessed,
ensuring linear scaling without performance degradation or cross-version interference.
Experiments were run on our own server with
4 CPU cores (11th Gen Intel Core i7),
32 GB of RAM, 1 TB SSD, and
a GeForce RTX 3070 GPU with 8GB GDDR6.

\begin{figure}
      \includegraphics[width=0.9\columnwidth]{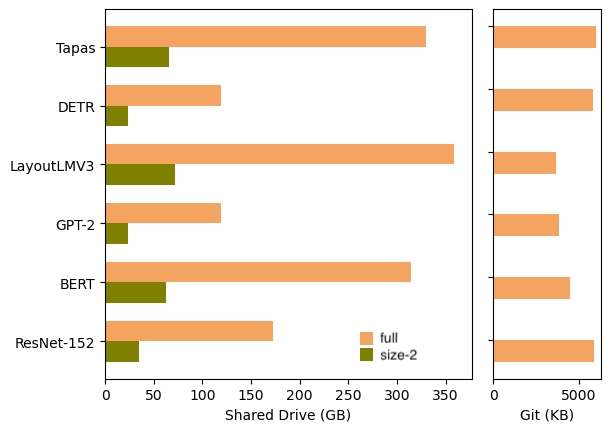}
      \caption{Storage Requirements}
      \label{fig:storageFootprint}
\end{figure}

\subsection{Efficient Linear Scaling across Versions}
To assess FlorDB's ability to manage multiversion hindsight logging without cross-version interference, we evaluated its performance across different model architectures.
For each model architecture, we measured the runtime for the following modes of execution (as defined in \Cref{sec:costPredictor}):

\begin{itemize}
      \item \textbf{record}: First run of model training; generates checkpoints.
      \item \textbf{replay-prefix}: query runs a \texttt{prefix scan} over all versions.
      \item \textbf{replay-suffix}: query runs a \texttt{suffix scan} over all versions.
      \item \textbf{replay-validation}: query runs a \texttt{validation scan} over all versions.
\end{itemize}
Our findings, depicted in \Cref{fig:linePlots}, reveal that FlorDB's time complexity scales linearly with an increasing number of versions, as demonstrated by the log-scale on the y-axis. This evidence of linear scalability, devoid of interaction effects or incremental overhead, validates FlorDB's ability to handle both large models and big datasets.

\subsubsection{Interactive Response Times}
In evaluating the impact of different replay operations on system response times (see orange lines in \Cref{fig:linePlots}),
the replay-prefix operation maintains the fastest processing times, consistently delivering results in under 10 seconds. Even as the number of versions increases, the processing time for replay-prefix remains the most stable, ensuring a responsive user experience. In contrast, replay-suffix and replay-validation operations exhibit longer processing times as the version count rises, with replay-validation times increasing more steeply. One factor is the amount of time it takes to evaluate the model. While replay-suffix may still offer interactive speeds at lower version counts, replay-validation quickly surpasses the interactive threshold:
in scenarios where immediate feedback is crucial, users can refine the selectivity of their queries
for faster response times (as discussed in \Cref{sec:costPredictor}).

\subsubsection{Storage Requirements}
Despite frequent commits across numerous versions, Git repositories remain remarkably compact, typically under 5MB (see right pane of \Cref{fig:storageFootprint}).  In contrast, checkpoints for large models stored on shared storage can easily consume hundreds of gigabytes (see orange bars in the left pane).
To address potential local storage depletion, FlorDB's default approach is to spool least-recently-used checkpoints to a cloud object store like S3\footnote{Storing 100 GB of data in S3 costs approximately \$2.30 a month.}. If this is not feasible, FlorDB may retain a subset of checkpoint per version---as few as just two (see green bars in \Cref{fig:storageFootprint})---evicting the rest. FlorDB~retains checkpoints that are evenly spaced across iterations. This ensures balanced work partitions and avoids stragglers that would hinder replay parallelism.
Importantly, even retaining just two checkpoints (one mid-run, one at completion) offers significant benefits:
i) the final checkpoint enables fast restores for replay-prefix and replay-suffix queries, and the mid-run checkpoint enables 2x faster replay execution compared to the original run.

\begin{figure}
      \centering
      \includegraphics[width=0.72\columnwidth]{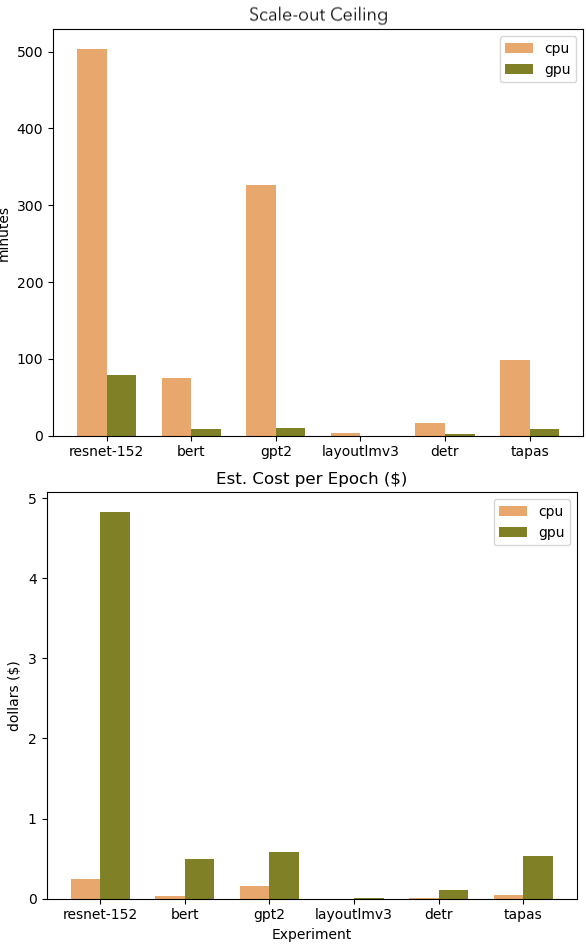}
      \caption{Comparison of runtime (top) vs rental cost (bot).}
      \label{fig:scaleout}
\end{figure}

\subsection{Speed-up through Selectivity \& Scale-out}
\label{sec:evalselectivity}
As discussed in the previous section,
multiversion hindsight logging queries
may be long-running tasks. In order to maintain interactive response times,
it may be desirable to refine the query to be more selective (Section~\ref{sec:costPredictor}),
or to allocate more resources (e.g., GPUs) and run the query in parallel.
In some cases, a developer may be willing to pay the cost of generating extensive hindsight log messages from the inner loop of their training script. In this case each iteration of the inner loop must be replayed from checkpoint, and logs generated; a compute- and data-intensive task. In these cases, parallelism comes into play: because each loop iteration is based on a separate checkpoint, we can replay as many loop iterations as possible at once with embarrassing parallelism. We refer the reader to the experimental results in our earlier work measuring this effect~\cite{garcia2020hindsight}.
To address the potential for long query times and the desire for selectivity or resource scaling, let's now examine the bottlenecks involved and the cost/performance trade-offs associated with resource use.

\subsubsection{Bottleneck Analysis --- length-1 range scan}
\Cref{fig:scaleout} (top) depicts the runtime of a range scan over a single epoch.
Assuming \emph{unbounded} resources (e.g. in an idealized cloud),
every version would be replayed in its own machine.
In this ideal scenario, a query's bottleneck would be
the time it takes for a single Flor replay operation
to finish. Replay-prefix, replay-suffix, and replay-validation
would each finish in about 100 seconds or less (\Cref{fig:linePlots}).
In contrast, the cost of a range scan, even for a single epoch, is higher. This operation requires stepping into the nested training loop, performing a forward pass over the neural network, and back-propagating gradients. Consequently, the range scan emerges as a bottleneck operator. In \Cref{fig:scaleout} we evaluate its runtime on a single epoch; running times for more epochs or versions can be linearly extrapolated.
This is due to the ideal parallelism of replay-from-checkpoint (confirmed by prior studies~\cite{garcia2020hindsight}) and the embarrassing parallelism of replay
across versions.

\subsubsection{Cost and Performance Trade-offs}
Our bottleneck analysis assumed an ideal of unbounded resources.
While parallelism can significantly accelerate computation, it comes with increased resource consumption and financial costs. \Cref{fig:scaleout} (bottom) compares the estimated cost per epoch across CPU and GPU platforms. Costs are based on AWS EC2 instance rental fees, with and without GPUs, factoring in execution time.  GPUs provide order-of-magnitude superior computational speed but incur order-of-magnitude higher expenses.
CPUs have the added benefit that they can be elastically allocated \emph{en masse}. This can improve throughput across many versions or epochs, but
as the bottleneck analysis indicates, there are limits to response times when range scans are used.
\section{Related Work}
ModelDB \cite{vartak2016modeldb} is a system designed to store, version, and manage ML models effectively. 
Similar systems include Weights \& Biases~\cite{wandb}, VisTrails~\cite{bavoil2005vistrails}, and others~\cite{vartak2018mistique, breck2019data, kumar2017data}.
ModelDB allows users to log complete model metadata, including hyper-parameters, data splits, evaluation metrics, and the final model file. Moreover, it provides the ability to query over this metadata, making it an effective tool for analysis and comparison of different ML models and experiment runs.
However, the focus of ModelDB is mainly on managing the straightforward metadata of deployed models, 
and it does not provide any features for hindsight logging. 
ModelDB's focus on metadata is orthogonal to FlorDB's multiversion hindsight logging facilities; 
the two systems could be used in concert or individually.

MLFlow \cite{zaharia2018accelerating} 
is an open-source platform that helps manage the end-to-end machine learning lifecycle, including experimentation, reproducibility, and deployment. 
It provides functionalities to log parameters, versioned code, metrics, and output artifacts from each run and later query them. Its modular design allows it to be used with any existing ML library and development process.
Similar systems include TFX~\cite{baylor2017tfx}, Airflow~\cite{ApacheAirflow}, Helix~\cite{xin12helix}, and others~\cite{anil2020apache, GoogleVertexAI}.
Like ModelDB, MLFlow has no support for hindsight logging or versioned log management. However, also like ModelDB, MLFlow and FlorDB can be fruitfully used together or separately.

FlorDB's query model builds upon the notion of Acquisitional Query Processing (AQP)~\cite{madden2005tinydb}. Traditional query processing assumes that data is pre-stored and ready for querying; AQP introduced the concept of efficiently acquiring data as a part of the query process. The idea is particularly useful for applications like sensor networks, where querying can be expensive in terms of energy or computational resources. 
FlorDB's multiversion hindsight logging also uses a form of AQP, where queries are not just made over pre-stored data, but also consider acquiring data through experiment replay.
Unlike the original work on AQP, our acquisition task---i.e., multiversion hindsight logging---raises unique technical challenges of its own, forming the bulk of our work.

R3 \cite{li2023r3}, with its record-replay-retroaction mechanism, is primarily focused on database queries and transactions, capturing and replaying states at the database level. This approach is efficient for debugging with low-overhead recording and storage-efficient replay.
While R3 aims to enhance the debugging and testing capabilities of database-backed applications, 
FlorDB addresses the specific needs of machine learning model development and analysis with its hindsight logging and query processing features.
\section{Conclusion}\label{sec:conclusion_future_work}
This paper introduces FlorDB, 
a system designed to address the unique challenges faced by machine learning engineers (MLEs) 
in managing the iterative, data-rich model development process. 
FlorDB's approach to multiversion hindsight logging, a record-replay technique, 
allows MLEs to add logging statements post-hoc, 
thereby enabling MLEs to ``query the past.''
FlorDB's key contributions include a unified relational model for querying log results, 
automatic propagation of new logging statements across versions 
as part of acquisitional query processing, 
and a highly-accurate empirical cost predictor for replay query refinement. 
These features streamline the ML experimentation process, enabling more efficient analysis and 
faster iteration. We evaluate the system's performance across various computer vision and NLP benchmarks, 
demonstrating the scalability of FlorDB across multiple versions.

\begin{acks}
      This work is supported in part by National
      Science Foundation CISE Expeditions Award CCF-1730628,
      IIS-1955488, IIS-2027575, DOE award DE-SC0016260, ARO
      award W911NF2110339, and ONR award N00014-21-1-2724.
      Rolando Garcia is supported in part by a grant from Dell Inc.
      We also thank Robert Lincourt, Matei Zaharia, and Bobby Yan
      for helpful comments and feedback.
\end{acks}

\bibliographystyle{ACM-Reference-Format}
\bibliography{main}

\end{document}